\documentclass[twocolumn]{aastex631}

\begin{document}

\title{Strong NIR emission following the long duration GRB\,211211A: Dust heating as an alternative to a kilonova}

\author[0000-0002-9038-5877]{Eli Waxman}
\affiliation{Dept. of Particle Phys. \& Astrophys., Weizmann Institute of Science, Rehovot 76100, Israel}

\author{Eran O. Ofek}
\affiliation{Dept. of Particle Phys. \& Astrophys., Weizmann Institute of Science, Rehovot 76100, Israel}

\author{Doron Kushnir}
\affiliation{Dept. of Particle Phys. \& Astrophys., Weizmann Institute of Science, Rehovot 76100, Israel}

\begin{abstract}

The prolonged near infrared (NIR) emission observed following the long duration GRB\,211211A is inconsistent with afterglow emission from the shock driven into the circum-stellar medium (CSM), and with emission from a possible underlying supernova. It has therefore been suggested that the observed NIR flux is the signature of a kilonova -- a radioactive ejecta that is similar to the outcome of the binary neutron star merger GW170817. We propose here an alternative plausible explanation. We show that the NIR flux is consistent with thermal emission from dust, heated by UV and soft X-ray radiation produced by the interaction of the GRB jet plasma with the CSM. This NIR emission was predicted by Waxman \& Draine for GRBs residing near or withing massive molecular clouds.
The dust NIR emission scenario is consistent with a GRB at $z\lesssim1$. Inspection of the environment of GRB\,211211A suggests that there are at least two host-galaxy candidates, one at $z=0.076$ and the other at $z=0.459$. The $z=0.459$ possibility is also consistent with the non-detection of a supernova signature in the light curve of the GRB afterglow, and with a typical GRB $\gamma$-ray energy for the fluence of GRB\,211211A.

\end{abstract}


\section{Introduction} \label{sec:intro}

\citet{2022lGRB-KN-Rastinejad} reported the detection of near infrared (NIR) emission, $f_\nu\approx 2\,\mu$Jy at $2.2\,\mu$m ($K$-band), lasting for $\sim10$\,d following the long (50~s) duration gamma-ray burst GRB\,211211A.
The magnitude and duration of the NIR emission are both too large to be consistent with the {\it afterglow} emission, produced by the shock-wave driven by the GRB jet into the circum-stellar medium (CSM), given the measured X-ray afterglow flux. The large ratio of NIR to shorter wave-length flux, $f_{\nu,K}/f_{\nu, i}\approx 30$ at $t\approx 5$\,d, is inconsistent with non-thermal afterglow shock emission, and challenging for an explanation of the IR flux as due to an underlying supernova.
It has therefore been suggested \citep{2022lGRB-KN-Rastinejad,2022lGRB-KN-Troja,2022lGRB-KN-Yang} that the NIR emission is due to a {\it kilonova} -- a radioactive ejecta that is similar to the outcome of the binary neutron star merger GW170817 \citep{Abbott17PhRvL}. The opacity of kilonova ejetca may be large due to the presence of significant amounts of Lanthanide elements \citep{BarnesKasen13}, leading to strong emission in the NIR \citep[see e.g.][for review]{2016FernandezMetzgerRev}.  We propose here an alternative plausible explanation for the NIR emission.

NIR emission, $L_{\rm NIR}\approx 10^{41}$\,erg\,s$^{-1}$ lasting for $\approx20$\,d, was predicted to follow energetic GRBs, which reside within or near massive molecular clouds \citep[][hereafter WD00]{WaxmanDraine00}. At the onset of the deceleration of the GRB jet plasma by the CSM, strong optical-UV-X-ray emission is produced by the forward shock driven into the CSM and by the reverse shock driven back into, and decelerating, the jet plasma. The resulting optical-UV luminosity, $\approx 10^{49}$\,erg\,s$^{-1}$ over $\approx10$\,s, can destroy dust by sublimation out to large radius, $\approx10$\,pc, within the beam of the jet. It was shown in WD00 that if the (residual) optical depth of the cloud's dust beyond this distance is significant, with optical depth of $0.2\lesssim \tau_V\lesssim 2$ in the visible, most of the optical-UV energy is converted to NIR emission peaking at the $K$-band. We show here that the NIR emission observed following GRB 211211A is consistent with the predicted heated dust emission, for cloud parameters (mass and radius) that are common to molecular clouds (see discussion in \S~\ref{sec:summary}).

The detection of the prompt optical-UV flash is challenging due to its short, $\sim10$\,s duration. A prominent example is GRB 990123, where contemporaneous gamma-ray and optical observations revealed a 9th mag optical flux at 10\,s \citep[][]{1999Natur.398..400A}.
Other examples include GRB\,080319B and GRB\,061007 (see e.g., \citealt{Kann+2010ApJ_GRB_SwiftLC}).
\citet{2014ApJ...781...37P} considered the optical-UV afterglow detected at early times, $t\gtrsim100$\,s by {\it Swift}, and found that a significant fraction of $\gamma$-ray luminous bursts show $L\sim10^{49}$\,erg\,s$^{-1}$ at $U$-band at $10^2-10^3$\,s. Direct evidence for dust destruction, through the detection of a significant reduction in extinction during the prompt flash, is even more challenging.
Such an evidence has been obtained by \citet{2014MNRAS.440.1810M}, who found a a significant red-to-blue color evolution of the optical light at times of $15-200$\,s following the burst trigger.
Indirect evidence for dust destruction, through the detection of the long term NIR emission that is expected for $0.2\lesssim \tau_V\lesssim 2$, may have been observed in a few cases (e.g. GRB\,970228 and GRB\,980326, see WD00).

In \S~\ref{sec:dust} we briefly explain the dust NIR emission model. We do not repeat the derivation of WD00, but provide a simple explanation of the main results. In \S~\ref{sec:GRBA} we show that the NIR emission of GRB\,211211A is consistent with heated dust emission, and in \S~\ref{sec:earlyOX} we show that early ($\sim100$~s) X-ray and UV observations are consistent with this scenario. \citet{2022lGRB-KN-Rastinejad} associate the GRB with a galaxy at a redshift $z=0.076$, corresponding to a distance of 350\,Mpc. However, since the probability of a chance coincidence between the GRB and the galaxy is significant, of order a few percent, we consider also the possibility that the GRB host galaxy is significantly more distant.
We focus on the explanation of the NIR emission since it is the key characteristic that distinguishes the afterglow of GRB\,211211A from those of other GRBs.
Other afterglow characteristics are not uncommon, as we explain in \S~\ref{sec:AG}. In \S~\ref{sec:host_SN} we discuss the constraints imposed by the optical-NIR observations on a possible host galaxy and a possible underlying supernova.
We note that the measured $K$- to $i$-band specific flux ratio, $\approx30$  at $5$\,d, is much larger than that observed, at a similar time, in the kilonova emission following the binary neutron star merger detected by LIGO, GW170817 \citep{Abbott17PhRvL}, where the flux ratio is $\approx7$ (e.g., \citealt{2018WOKG}), and is difficult to reconcile with existing models for any radioactive ejecta. Our conclusions are summarized in \S~\ref{sec:summary}.

Before addressing the unusual IR properties of GRB\,211211A, a comment is in place regarding its $\gamma$-ray properties. The distance-independent $\gamma$-ray properties of GRB211211A, \{hardness ratio, duration, fluence, minimum variability time, "effective amplitude parameter"\}, are all well within the ranges characteristic of long GRBs, and inconsistent with those of short GRBs: GRB211211A is located at the center of the long GRB population, and is well separated from the short GRB population, in the \{hardness ratio/duration, fluence/minimum variability time, "effective amplitude parameter"/duration\} planes; it is well separated from the short GRB population, and close to the high fluence end of the long GRB population, in the fluence/variability plane (Fig.~1 of \citet{2022lGRB-KN-Rastinejad}, Extended Data Fig.~2 of \citet{2022lGRB-KN-Troja}, Fig.~2 of \citet{2022lGRB-KN-Yang}). In the two distance-dependent relations planes, \{peak luminosity/temporal lag, spectral peak energy/isotropic $\gamma$-ray energy\} GRB211211A lies outside the bands occupied by the long GRB and short GRB populations for an assumed redshift of $z=0.08$ (Extended Data Fig.~2 of \citet{2022lGRB-KN-Troja}, Fig.~2 and Table~1 of \citet{2022lGRB-KN-Yang}), and well within the bands occupied by the long GRB population for an assumed redshit of $z\approx0.5$. Thus, while the $\gamma$-ray properties of GRB211211A are inconsistent with those characterizing short GRBs, and its distance-dependent properties are a-typical for long duration GRBs assuming a $z=0.08$ origin, its properties are typical for long duration GRBs assuming $z\sim0.5$.

\section{Dust heating interpretation of the NIR emission of GRB211211A}
\label{sec:dust_GRBA}

We discuss in this section a plausible explanation of the strong prolonged NIR emission, which lasts for $\sim10$\,d with $f_\nu\approx 2\,\mu$Jy at $2.2\,\mu$m ($K$-band). The very large ratio of IR to shorter wave-length optical flux, $f_{\nu,K}/f_{\nu, i}\approx 30$ at $t\approx 5$\,d (where $i$-band corresponds to $\approx0.75\,\mu$m), cannot be explained by synchrotron emission from electrons accelerated by the expanding collisionless afterglow shock. This is due to the fact that shock acceleration leads to an electron distribution $dn_e/dE_e\propto E_e^{-p}$ with a power-law index $p\approx2$, yielding $f_\nu\propto \nu^{-s}$ with $s$ in the range of $(p-2)/2$ to $p/2$, i.e. $\approx 0.5-1$ \citep[see][for reviews of afterglow theory]{2004RvMP...76.1143P,2006RPPh...69.2259M,2006PPCF...48B.137W}. The large NIR to optical flux ratio is also inconsistent with emission from an underlying supernova (see Section~\ref{sec:SN}). It therefore requires another source of radiation, which we show in \S~\ref{sec:dust} and \S~\ref{sec:GRBA} to be consistent with thermal dust emission.

In \S~\ref{sec:earlyOX} we show that early ($\sim100$~s) X-ray and UV observations are consistent with the dust emission scenario, and that the UV flux measurements at $\sim100$~s set relevant constraints on model parameters (in particular on the CSM density). As mentioned in the introduction, and discussed in \S\ref{sec:AG}, the afterglow ($\gtrsim1$~hr) behavior of GRB211211A at the X-ray and optical bands, excluding the NIR band, is not uncommon. Furthermore, we explain in \S\ref{sec:AG} that, due to large uncertainties in afterglow modelling, this afterglow data do not enable one to derive useful constraints on the parameters of the jet and CSM. We therefore do not discuss the optical-X-ray afterglow in detail.

\subsection{Heated dust emission}
\label{sec:dust}

We consider a highly relativistic jet, with isotropic equivalent energy $E$ and opening angle $\theta$, propagating into a uniform medium of number-density $n$ (note that $n$ is the number density at the $\sim10^{17}$\,cm vicinity of the source, which may differ greatly from the density at larger distances). The expanding plasma drives a highly relativistic forward shock into the surrounding CSM, and the elevated CSM pressure drives a reverse shock into the jet plasma. After the reverse shock crosses the jet plasma, most of the energy is transferred to the CSM and the flow approaches a self-similar behavior \citep{BM76}. The forward shock that continues to propagate into the CSM produces the afterglow by synchrotron emission of shock accelerated electrons. We are interested here at the early transition stage, where the reverse shock is going through the jet plasma.

The (observed) duration $\Delta t_{\rm tr}^{\rm obs}$ of the transition stage (see Eq.~2 of WD00) is the larger of the duration over which most of the energy (carried by the jet) is emitted from the source, which is approximately given by the (observed) duration of the GRB $\gamma$-ray emission $\Delta t_{\rm GRB}^{\rm obs}$, and the (observed) time $\Delta t_\Gamma$ at which the Lorentz factor of the self-similar flow described by the \citet{BM76} solution equals the original ejecta Lorentz factor $\Gamma_i$,
\begin{eqnarray}\label{eq:t_Gamma}
  \Delta t_{\rm tr}^{\rm obs} &=& \max\left[\Delta t_{\rm GRB}^{\rm obs},\Delta t_\Gamma^{\rm obs}\right], \nonumber\\
  \Delta t_\Gamma^{\rm obs} &=& 8.4\frac{1+z}{1.4}\left(\frac{E/10^{53}\rm erg}{n/1\rm cm^{-3}}\right)^{1/3}\left(\frac{\Gamma_i}{300}\right)^{-8/3}{\rm s}.
\end{eqnarray}
For typical long GRB parameters, $\Delta t_{\rm GRB}\approx\Delta t_\Gamma\approx10$~s.
Towards the end of this stage,
most of the energy is carried by the thermal energy of the shock heated CSM/jet plasma. Denoting the fraction of thermal energy carried by shock accelerated electrons by $\epsilon_e$, the energy that may be radiated at this stage is $\approx \epsilon_e E$. For typical GRB parameters, the reverse shock produces optical-X-ray radiation, while the forward shock produces soft-hard X-rays.

WD00 considered only the energy radiated within the energy range of $1-7.5$\,eV as available for dust destruction and heating, due to the fact that in dense clouds most of the radiation energy carried by photons of energy $>13.6$\,eV goes to ionization, and most of the energy in the $7.5-13.6$\,eV is absorbed by H$_2$. For a flat electron energy distribution, $E_e^{2}dn_e/dE_e\propto E_e^{0}$, as expected for shock acceleration, the fraction of energy carried by electrons radiating at this energy range is approximately $\log(7.5)/\log(E_{\rm max}/E_{\rm min})\approx 0.1$ (assuming $E_{\rm max}/E_{\rm min}\approx 10^{10}$).

The fraction of electron energy lost to radiation depends on the ratio between their cooling time and the expansion time of the plasma. The cooling frequency, i.e. the synchrotron radiation frequency of electrons with cooling time comparable to the expansion time, is given by (see Eq.~7 of WD00)
\begin{eqnarray}\label{eq:nu_c} 
  \nu_c \approx 7&\times&10^{16}\left(\frac{\epsilon_B}{0.01}\right)^{-3/2}\left(\frac{n}{1\rm cm^{-3}}\right)^{-1} \nonumber\\
   &\times& \left(\frac{1.4}{1+z}\frac{E}{10^{53}\rm erg}\frac{\Delta t_{\rm tr}^{\rm obs}}{10~\rm s}\right)^{-1/2}{\rm Hz},
\end{eqnarray}
where $\epsilon_B$ is the fraction of post shock thermal energy carried by magnetic fields.
Electrons radiating at higher frequencies lose all their energy to radiation, while electrons radiating at lower energy lose a fraction $(\nu/\nu_c)^{1/2}$ of their energy to radiation (and the rest to adiabatic expansion). WD00 adopted a value of $n=1\,{\rm cm^{-3}}$, for which electrons in the range $1-7.5$\,eV lose $\approx 10\%$ of their energy to radiation, implying that the radiated energy available for dust heating is $\approx 0.01 \epsilon_e E$ (see Eq.~6 of WD00).
For bright bursts, or lower mass clouds, the energy carried by ionizing radiation at 7.5-50~eV may be large compared to that required to fully ionize the gas cloud. In this case, a significant fraction of the 7.5-50~eV radiation will contribute to dust heating, increasing the luminosity available for dust heating by a factor of a few (see WD00 and \S~\ref{sec:GRBA}). Finally, the flux of higher energy, $>0.1$~keV photons, for which absorption is dominated by ionization of "metals" (i.e. atoms heavier than He), may further contribute to dust heating \citep[e.g.][]{2001FruchterDust}: For bright bursts the contribution of $\sim1$~keV photons is similar to that of the non-ionizing $1-7.5$~eV photons, while the contribution of higher energy, $h\nu>10$~keV, photons is small- see a short discussion in appendix \ref{sec:appendix}.

For the current discussion, we denote by $f_d$ the fraction of the total electron energy, $\epsilon_e E$, lost to radiation in the spectral band available for dust destruction. The energy and luminosity of this radiation are thus given by
\begin{equation}
\label{eq:Ld}
    E_d\equiv f_d\epsilon_e E,\quad L_d=\frac{(1+z) E_d}{\Delta t_{\rm tr}^{\rm obs}}.
\end{equation}
Here, $L_{d}$ is the destruction luminosity (associated with $E_{d}$)
and $\Delta t_{\rm tr}^{\rm obs}/(1+z)$ is the pulse duration at the host galaxy frame. Based on the discussion of the preceding paragraph, we expect $f_d\gtrsim10^{-2}$.

The rapid expansion of the shock heated jet plasma, following its crossing by the reverse shock, rapidly reduces the characteristic synchrotron emission frequency of the highest energy electrons, which did not lose all their energy to synchrotron emission during the prompt flash emission (i.e. electrons that initially produce radiation at $\nu\sim\nu_c$). The synchrotron emission frequency of the highest energy electrons drops below the optical band on a time scale of tens of seconds, leading to a strong suppression of the flux on this time scale \citep[e.g.][]{2000Kobayashi_RS}.

The optical-UV and soft, $\sim1$~keV, photon flash is capable of destroying dust out to a distance $R_d$, which depends on $L_d$ and on the properties of the dust. For the current discussion, we will present results normalized to dust parameters appropriate for graphite materials and grain size of $0.3\,\mu$m. For such grains, the temperature for full sublimation to occur over $\Delta t_{\rm tr}=10$\,s is (see Eq.~11 of WD00)
\begin{equation}
\label{eq:Tc}
    T_c\approx 2700\,{\rm K},
\end{equation}
and the destruction radius is (see Eq.~17 of WD00)
\begin{equation}
\label{eq:Rd}
    R_d\approx 1.3\times10^{19}\left(\frac{T_c}{2700\,\rm K}\right)^{-2}L^{1/2}_{d,49}\,\rm cm,
\end{equation}
where $L_d=10^{49}L_{d,49}$\,erg\,s$^{-1}$.

Dust is destroyed out to distance $R_d$ within the GRB jet beam. Dust at somewhat larger distance is heated to $T\approx T_c$, converting the incoming UV radiation to NIR radiation. If the optical depth of the cloud beyond $R_d$ is large for UV photons, and small for NIR photons, a significant fraction of the {\it destructive flash} energy, $E_d$, will be radiated in the NIR. For absorption efficiency $Q_\nu\propto \nu$, large optical depth in the UV and small optical depth in the NIR may be written as $0.2\lesssim \tau_V\lesssim 2$. The total energy radiated is $0.5\theta^2 f_d\epsilon_e E$ (i.e. the true, rather than isotropic equivalent energy), and the observed duration of the pulse (without redshift correction) is $0.5\theta^2 R_d/c$, thus the NIR luminosity is
\begin{equation}
\label{eq:L_IR}
    L^{\rm Dust}_{\rm NIR}=\frac{f_d\epsilon_e E}{R_d/c}.
\end{equation}

\subsection{GRB 211211A}
\label{sec:GRBA}

Let us consider now the case of GRB 211211A. An estimate of $\epsilon_e E$, the (isotropic-equivalent) thermal energy carried by the fireball electrons at the onset of significant deceleration, may be obtained from the observed energy released in $\gamma$-rays. $\gamma$-rays are produced as part of the energy of the jet is converted, prior to the interaction with the CSM, to internal energy carried by electrons, and then radiated away as $\gamma$-rays. Denoting this fraction by $f_i$, we have $\epsilon_e E\approx f_i^{-1}E_\gamma$. The tight correlation between $\gamma$-ray energy and X-ray afterglow flux
implies that the efficiency of fireball energy conversion to $\gamma$-rays is not small, i.e. that $f_i$ is of order unity \citep{2001ApJ...547..922F,2016ApJ...824..127W}. Using the $\gamma$-ray fluence, $F_\gamma=5\times10^{-4}$\,erg\,cm$^{-2}$ \citep{2021GCN.31210....1M}, we have
\begin{equation}
\label{eq:Eg}
    \epsilon_e E\approx 1.8\times 10^{53}\frac{1.4d_{\rm 2 Gpc}^2}{1+z}\,{\rm erg}.
\end{equation}
Here, $d_L=2d_{\rm 2\,Gpc}$\,Gpc, corresponding to $z=0.4$.

Using equations~(\ref{eq:Ld}) and~(\ref{eq:Eg}) we obtain an estimate for the {\it sublimation flash} luminosity,
\begin{equation}
\label{eq:L_dA}
    L_d\approx 9.7\times10^{49} f_{d,-2}\left(\frac{\Delta t_{\rm tr}^{\rm obs}}{25~\rm s}\right)^{-1}d_{\rm 2 Gpc}^2\,{\rm erg\,s}^{-1},
\end{equation}
where $f_d=10^{-2}f_{d,-2}$ and we present numerical results with $\Delta t_{\rm tr}$ normalized to 25~s, a duration including both the main, $\sim10$~s, $>100$~keV $\gamma$-ray pulse during which $\approx80$~\% of the total fluence (10~keV-1~MeV) is observed, and the second smaller $\gamma$-ray pulse observed between 15~s and 25~s following the GRB onset. The results depend weakly on $\Delta t_{\rm tr}$.

The dust destruction distance may now be obtained from Equation~(\ref{eq:Rd}),
\begin{equation}
\label{eq:RdA}
    R_d\approx 4.0\times10^{19}\left(\frac{T_c}{2700\,\rm K}\right)^{-2} f_{d,-2}^{1/2}d_{\rm 2 Gpc}\,{\rm cm},
\end{equation}
and the dust NIR luminosity from Equtaion~(\ref{eq:L_IR}),
\begin{equation}
\label{eq:L_IRA}
    L^{\rm Dust}_{\rm NIR}\approx1.9\times10^{42}f_{d,-2}^{1/2}\frac{1.4d_{\rm 2 Gpc}}{1+z}\,{\rm erg\,s}^{-1}
\end{equation}
($R_d\propto \Delta t_{\rm tr}^{-1/2}$, $L^{\rm Dust}_{\rm NIR}\propto \Delta t_{\rm tr}^{1/2}$). For thermal emission, $\nu L_\nu$ peaks at $h\nu=4T$, corresponding to
\begin{equation}
\label{eq:lam_p}
    \lambda_{\rm peak}^{\rm Dust}=(1+z)\frac{hc}{4T_c}=1.7\frac{1+z}{1.4}\frac{2700\,{\rm K}}{T_c}\mu{\rm m}.
\end{equation}

We may now compare the predicted dust luminosity to the observed luminosity, which we estimate as $4\pi d_L^2(\nu f_\nu)_K$ with $f_{\nu,K}=2\,\mu$Jy\footnote{There are four epochs with $K$-band detection: $\approx4\,\mu$Jy at $t\approx4-5$\,d and $\approx1-1.5\,\mu$Jy at $t\approx7-8$\,d. We use a time averaged $2\,\mu$Jy.},
\begin{equation}
\label{eq:Lratio}
    \frac{L^{\rm Dust}_{\rm NIR}}{(\nu L_\nu^{\rm obs})_{K}}\approx
    1.0\left(\frac{T_c}{2700\,\rm K}\right)^{2}
    \frac{1.4}{(1+z)}
    \frac{f_{d,-2}^{1/2}}{d_{\rm 2\,Gpc}}.
\end{equation}
The observed NIR emission is thus consistent with a dust emission for a GRB at $z\approx0.5$. An accurate redshift determination is not possible based on Eq.~(\ref{eq:Lratio}) due to the uncertainties in $T_c$ and $f_d$ and in the details of the underlying model (e.g. gas and dust density and distribution, jet structure, electron spectrum).
However, dust emission is unlikely to be a viable explanation at $z>1$, where the ratio of the predicted dust luminosity to the observed NIR luminosity drops to $\approx0.1$ and the emission peak shifts to longer wavelengths.
Note that dust emission provides a consistent explanation of the observed NIR emission for $z=0.076$, where the predicted dust luminosity is larger than the observed NIR luminosity (this may easily be accounted for, e.g., with somewhat smaller $f_d$ and $T_c$).

Let us consider next the flux ratio $f_{\nu,K}/f_{\nu, i}\approx 30$ at $t\approx 5$\,d. Assuming that the $i$-band luminosity is part of the Wien tail of the IR emission, and neglecting possible significant residual extinction at $i$-band, the $K$-band frequency needs to be close to that where the thermal emission peaks (the peak in $\nu L_\nu$ is at $h\nu=4T$, and $T$ should satisfy $h\nu_K=3.4T$), implying $T\approx2400([1+z]/1.4)$\,K.
This is consistent with the expected temperature for the dust properties that we have adopted, Eq.~(\ref{eq:Tc}), for $z\approx 0.5$. However, somewhat lower or higher temperatures may be obtained for different dust properties, thus allowing a wide redshift range.
Moreover, the $K$- to $i$-band flux ratio need not necessarily agree with a Planck spectrum ratio. It may be smaller, implying a possible contribution in the $i$-band from a different source (e.g. underlying supernova). It may also be larger, implying possible residual extinction.

The duration of NIR emission is longer than 8\,d. Recalling that the dust emission extends for $t_{\rm NIR}\approx0.5(1+z)\theta^2 R_d/c$, this sets a lower limit to the jet opening angle,
\begin{equation}
\label{eq:thetaD}
    \theta>\theta^{\rm Dust}\approx 0.031\left(\frac{1.4t_{\rm NIR}/10\rm d} {(1+z)f_{d,-2}^{1/2}d_{\rm 2 Gpc}}\right)^{1/2}\frac{T_c}{2700\,\rm K}
\end{equation}
($\theta\propto \Delta t_{\rm tr}^{1/4}$). An upper limit to the $K$-band flux, that is much lower than the observed flux at 8\,d is obtained at 89\,d, implying $\theta\le0.1$. This is consistent with the jet opening angle inferred from the break time in the X-ray afterglow, $t_X\approx 0.3$\,d at which the decline of the X-ray flux steepens, assuming that the break is due to jet sideways spreading \citep[][]{Rhoads99,1999ApJ...519L..17S,2000ApJ...538..187L}
\begin{equation}
\label{eq:thetaX}
    \theta^{\rm X}\approx 0.053(t_X/0.3{\rm d})^{3/8} \left(\frac{(1+z)^2d_{\rm 2Gpc}^2}{1.4^2\epsilon_{e,-1}n_0}\right)^{-1/8}.
\end{equation}
Here, $n=10^0n_0$\,cm$^{-3}$ and $\epsilon_e=10^{-1}\epsilon_{e,-1}$.

Finally, let us consider the effects of ionizing radiation, at 7.5--50\,eV. The luminosity carried by these photons, $L_{\rm ion}$ is similar or somewhat larger than that of the 1-7.5\,eV photons, $L_d$ (due to the shorter cooling time of higher energy radiating electrons). The (isotropic equivalent) Hydrogen mass that the prompt flash is able to ionize is (see Eq.~19 of WD00)
\begin{equation}
\label{eq:Mion}
    M_{\rm ion}\approx \frac{L_{\rm ion}}{L_d}\frac{m_pf_d\epsilon_eE}{25\,\rm eV}
    =4\times10^4 f_{d,-2}\frac{L_{\rm ion}}{L_d}\frac{1.4d^2_{\rm 2 Gpc}}{1+z}M_\odot.
\end{equation}
Thus, clouds of mass $\lesssim10^5\,M_\odot$ will be completely ionized. In this case, part of the ionizing radiation will contribute also to dust heating. Note that the energy of the "destructive flash" is sufficient for sublimating the dust in a cloud of mass $\approx10^4M_{\rm ion}$, since the dust mass is typically 1\% of the Hydrogen mass, and the energy (per mass) required for sublimation is much smaller than that required for ionization (see WD00).

\subsection{Early X-ray and UV observations}
\label{sec:earlyOX}

Optical-soft X-ray observations are not available for GRB\,211211A over the tens of seconds time scale expected for the emission from the reverse shock. We note, however, that the characteristics of the $>10$~keV flux observed at $t\sim50$~s \citep{2022lGRB-KN-Gompertz} are consistent with the theoretical model used in WD00, as described in \S~\ref{sec:dust}: The expected flux of rapidly cooling electrons (i.e. at $\nu>\nu_c$) in the forward shock, $\nu f_\nu\approx F_\gamma/(2t\log(E_{\rm max}/E_{\rm min}))=2\times10^{-7}{\rm erg/cm^2 s}$ is consistent with the observed $>10$~keV flux; The observed frequency dependence of the flux, $\nu f_\nu\approx1(h\nu/10\,{\rm keV})^{1/2}{\rm erg/cm^2 s}$ up to $\approx30$~keV, implies that $h\nu_c<10$~keV and that the synchrotron emission energy of the lowest energy electrons in the forward shock is $\approx30$~keV, which in turn implies that the synchrotron emission energy of the lowest energy electrons in the reverse shock, which is smaller by a factor approximately given by the square of the bulk jet Lorentz factor (WD00), is $\sim1$~eV.

The UV fluxes observed at 100~s \citep{2022lGRB-KN-Troja} and 160~s \citep{2022lGRB-KN-Gompertz} following the GRB onset are much lower than the destructive flash luminosity derived in the preceding sub-section. This implies that the characteristic (peak) synchrotron emission frequency $\nu_{\rm max}(t)$ of the highest energy electrons in the expanding reverse-shock heated jet plasma dropped well below $10^{15}$~Hz on a time scale of $\sim100$~s. The highest energy electrons are those that at $t\simeq\Delta t_{\rm tr}$ radiate at $\nu_c$ (higher energy electrons lose all their energy to synchrotron emission), $\nu_{\rm max}(t=\Delta t_{\rm tr})=\nu_c$. The flux at $\nu>\nu_{\rm max}$ drops with $t$ faster than exponentially \citep[and hence was taken to vanish in the analsysis of][]{2000Kobayashi_RS}, since after the reverse shock crossing, at $t>\Delta t_{\rm tr}$, electrons are no longer being accelerated in the expanding plasma and there are no electrons present with characteristic synchrotorn emission frequency above $\nu_{\rm max}$ (the flux at $\nu>\nu_{\rm max}$ is the exponentially decaying tail of the emission by electrons for which the synchrotron emission peaks at $\nu=\nu_{\rm max}$). Nevertheless, the strong suppression required for our model to be consistent with observations does set relevant requirements on model parameters. 

The time dependence of $\nu_{\rm max}(t)$ is uncertain since it depends on the time dependence of the (proper) density $n$ and (source frame) Lorentz factor $\gamma$ of the expanding plasma, which in turn depend on the unknown jet and CSM density distributions, and on the relation between $n$ and the magnetic field and electron energy densities. The flow behind the forward shock approaches the self-similar Blandford-McKee solution on a time scale similar to $\Delta t_{\rm tr}$. Assuming that the evolution of $n$ and $\gamma$ follows the self-similar solution, for which $n\propto R^{-13/2}$ and $\gamma\propto R^{-7/2}$ where $R$ is the shock radius, and assuming "adiabatic" expansion with magnetic and electron pressure proportional to $n^{4/3}$, $B^2\propto n\gamma_e\propto n^{4/3}$ where $\gamma_e$ is the electrons' Lorentz factor in the plasma frame, yields $\nu_{\rm max}\propto t^{-73/48}\approx t^{-3/2}$ \citep[e.g.][]{2000Kobayashi_RS}. 
For this behavior, the flux at $\nu=\nu_c$ drops by a factor of $10^5$ at $t/\Delta t_{\rm tr}=5$ (the time dependence approximately follows $3(t/\Delta t_{\rm tr})^{-1}\exp[-(t/\Delta t_{\rm tr})^{3/2}]$).

The rapid suppression of the flux above $\nu_{\rm max}$ can readily account for the $\sim100$~s UV observations, provided that $\nu_c=\nu_{\rm max}(t=\Delta t_{\rm tr})$ is close to  $10^{15}$~Hz. Using the energy estimate of Eq.~(\ref{eq:Eg}) and $\Delta t_{\rm tr}=25$~s in Eq.~(\ref{eq:nu_c}) we have 
\begin{equation}
\label{eq:nucut}
    \nu_c\approx 10^{15}\frac{1+z}{1.4d_{\rm 2Gpc}}
    \left(\frac{\epsilon_B}{0.01}\right)^{-3/2}
    \left(\frac{\epsilon_e}{0.1}\right)^{1/2}
    \left(\frac{n}{10\rm cm^{-3}}\right)^{-1}{\rm Hz}.
\end{equation}
The low UV flux at 100~s therefore implies $n(\epsilon_B/0.01)^{3/2}\gtrsim10\,{\rm cm^{-3}}$.

\subsection{Optical and X-ray afterglow}
\label{sec:AG}

The X-ray light curve shows a break at $t\approx0.3$\,d, which may be attributed to the onset of jet sideways expansion \citep{Rhoads99,1999ApJ...519L..17S}, and an optical to X-ray specific flux ratio varying from $f_{\nu,\rm O}/f_{\nu,\rm X}\approx50$ before the break to $f_{\nu,\rm O}/f_{\nu,\rm X}\approx200$ after it. The increased optical to X-ray flux ratio corresponds to a delay of the break time in optical flux, $\simeq0.7$\,d compared to $\simeq0.3$\,d in the X-rays. A flux ratio of $f_{\nu,\rm O}/f_{\nu,\rm X}\approx50-200$ is consistent with synchrotron emission from a power-law distribution of electrons, with a cooling frequency between the optical and X-ray bands, $h\nu_c\approx0.1$\,keV.

The evolution of the flux ratio $f_{\nu,\rm O}/f_{\nu,\rm X}$ during the break transition implies that the break is {\it chromatic}, while jet breaks are generally expected to be achromatic due to the relatively slow evolution of $\nu_c$ that is typically expected in afterglow models. However, such chromatic breaks are common. \citet{2008SwiftJetBreaks} analyze 60 GRBs with well sampled X-ray and optical afterglows with candidate jet breaks, finding that none are consistent with the simple achromatic jet-break predictions, and that the break times in the optical band are systematically longer, $\simeq3$~d, than those in the X-rays, $\simeq0.3$~d.

The deviation of the GRB\,211211A optical afterglow flux from the prediction of a simple afterglow model assuming an a-chromatic break \citep[e.g. Fig.~2 of][]{2022lGRB-KN-Troja} is thus common in GRB afterglows, and does not unambiguously imply the presence of an additional source of radiation.
Various explanations have been suggested to account for the observed chromatic breaks \citep[see, e.g.,][]{2006PanaitescuChromaticBreaks}. In general, it should be noted that afterglow models depend on a large number of free parameters (including jet energy $E$ and opening angle, CSM density $n$, post-shock energy fractions carried by electrons and magnetic fields, $\epsilon_e$ and $\epsilon_B$, electron spectral index $p$). Observations typically do not contain sufficient information for an accurate determination of these parameters, some of which (like $n$ and $\epsilon_B$) may vary by orders of magnitude. In the analysis of \citet{2022lGRB-KN-Rastinejad} of the afterglow of GRB 211211A, for example, the uncertainties in the inferred values of $E$, $n$ and $\epsilon_B$ are two, four, and three orders of magnitude, respectively. Moreover, afterglow models are based on simplifying assumptions regarding key underlying model components, including the (unknown) lateral and angular structure of the jet and of the CSM density distribution, and regarding possible time evolution of the micro-physical parameters ($\epsilon_e$, $\epsilon_B$, $p$), due to the evolving parameters (velocity, density) of the shock (it is typically assumed that these parameters are time independent). Deviations form this simplified description will lead to significant deviations from the model predictions.

\section{Host galaxy and underlying supernova} \label{sec:host_SN}

\subsection{Underlying SN}
\label{sec:SN}

The measured $K$- to $i$-band specific flux ratio, $\approx30$  at $5$\,d, is hard to reconcile with existing models for {\it any} radioactive ejecta. First note that this ratio is much larger than that observed at a similar time in the kilonova emission following the binary neutron star merger detected by LIGO, GW170817 \citep{Abbott17PhRvL}, where the flux ratio is $\approx7$ (e.g., \citealt{2018WOKG}). A similar $\lesssim10$ ratio is also obtained for the best fit kilonova models of \citet{2022lGRB-KN-Rastinejad}, despite a large freedom to fit many kilonova components.

Other known supernova types are also unable to explain such a ratio at $5$\,d \citep[see some examples in][]{2022lGRB-KN-Rastinejad}. \cite{2022lGRB-KN-Rastinejad} suggested that such a ratio can be explained by $100\%$ $^{56}$Ni ejecta with high, $\sim0.4c$, velocity ejecta. However, at $5\,\rm{d}$ such ejecta would be optically thin, where the fitting model of \cite{2022lGRB-KN-Rastinejad} is not applicable.
It is more likely that in this case the emission would be similar to the nebular phase of Type Ia SN or stripped-envelope SN types, where high $K$- to $i$-band specific flux ratio is not obtained.

While long/intermediate-duration GRBs without apparent supernovae were seen before (e.g., \citealt{GalYam2006b}; \citealt{Ofek2006GRB}), it is still possible that a supernova accompanied GRB 211211A but was too dim to be detected. Examining various GRB-associated SNe, \citet{2022lGRB-KN-Rastinejad} find that GRB 211211A is required to reside at $z\gtrsim0.5$ for the SN to go undetected. \citet{2022lGRB-KN-Troja} show that bright GRB-associated SNe, like SN\,1998bw and SN\,2006aj, would require larger distances, $z=0.8$ and $z=0.65$ respectively. However, modest extinction may allow even bright SNe to go undetected at $z\gtrsim0.5$- SN\,2006aj, for example, would be undetectable for host extinction of $E(B-V)\sim0.1$. Such reddening values are typical for GRBs \citep[][and references therein]{Covino13,Bolmer18}, and cannot be excluded for GRB 211211A (see \S~\ref{sec:summary}).

\subsection{Host}
\label{sec:host}

\begin{figure}
\includegraphics[width=8cm]{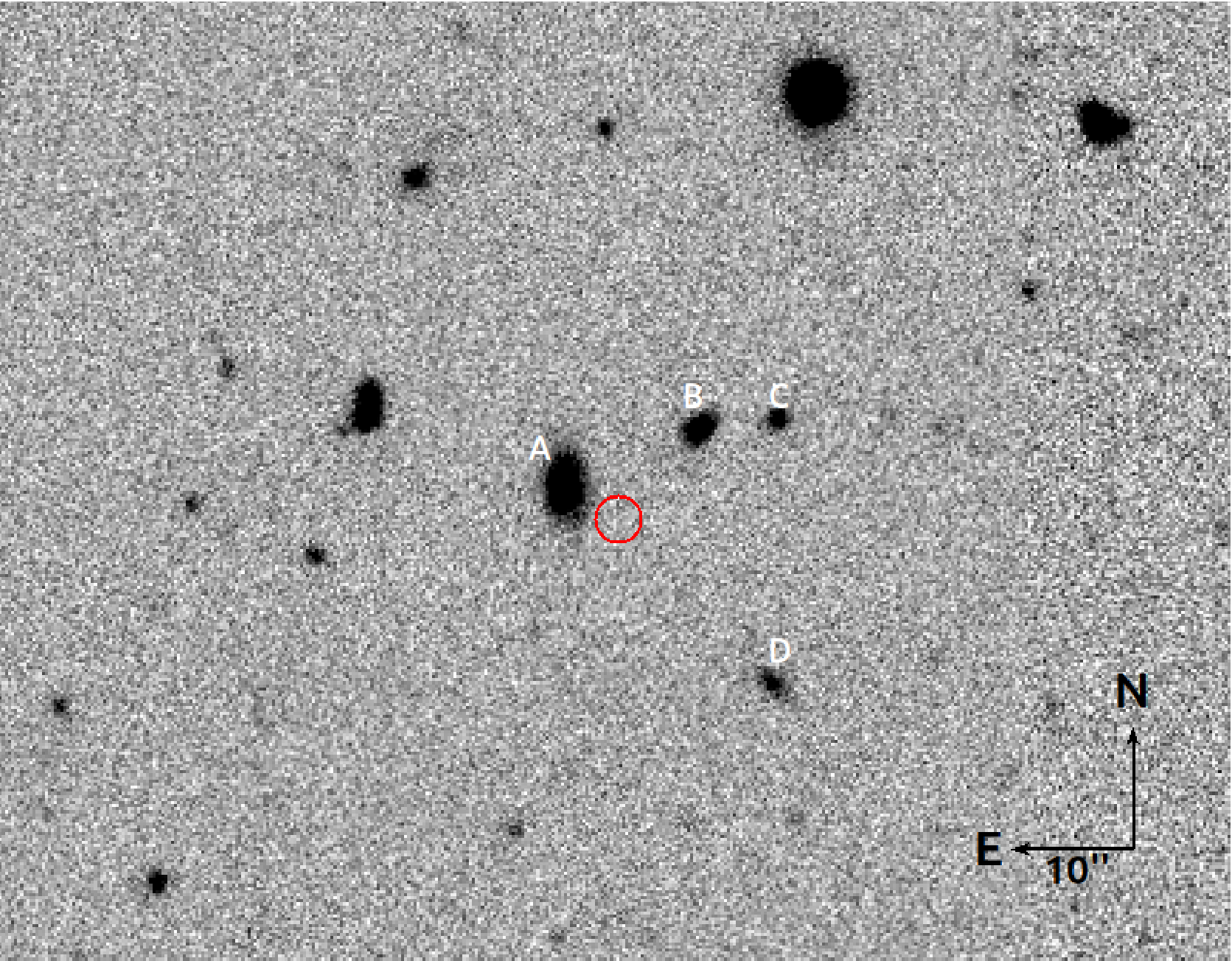}
\caption{DECaLS $r$-band caodd image of the field of GRB\,211211A.
The red circle marks the position of GRB\,211211A.
The properties of galaxies A to D are listed in Table~\ref{tab:Gals}.
The distances were calculated using tools presented in \cite{Ofek2014ascl.soft_MAAT}.
\label{fig:GRB211211A_DECaLS_r}}
\end{figure}

Inspection of the DECaLS (\citealt{Dey+2019AJ_DECaLS_summary}) and SDSS (\citealt{York+2000AJ_SDSS}) images of the GRB region
reveals a few spatially nearby galaxies.
Figure~\ref{fig:GRB211211A_DECaLS_r} shows the DECaLS DR9 $r$-band coadd image
around the GRB location, where the nearest four detectable galaxies are marked by $A$ to $D$.
Some fainter objects are also visible in the HST image presented in \cite{2022lGRB-KN-Rastinejad}.
Table~\ref{tab:Gals} list the DECaLS and SDSS measured properties of these galaxies.

\begin{deluxetable*}{lllll}
\tablecolumns{5}
\tablewidth{0pt}
\tablecaption{Properties of galaxies in the vicinity of GRB\,211211A}
\tablehead{
\colhead{Property}     &
\colhead{Gal. $A$}    &
\colhead{Gal. $B$}    &
\colhead{Gal. $C$}    &
\colhead{Gal. $D$} \\
\colhead{}    &
\colhead{}    &
\colhead{}    &
\colhead{}    &
\colhead{}
}
\startdata
RA                 & 14:09:10.47   & 14:09:09.60        & 14:09:09.09    & 14:09:09.13   \\
Dec                & $+$27:53:20.8 & $+$27:53:25.8      & $+$27:53:26.9  & $+$27:53:04.3   \\
g [mag]            & 19.99         & 22.3               & 22.5           & 23.4        \\
r [mag]            & 19.53         & 20.9               & 21.7           & 22.1        \\
z                  & 0.076         & $0.4587\pm0.0002$  &                &                \\
SDSS photo z       &               &                    & $0.45\pm0.13$  & $0.60\pm0.07$  \\
DECaLS photo z     &               &                    & $0.50\pm0.08$  & $0.53\pm0.07$  \\
abs. r [mag]       & $-18.1$       & $-21.1$            &                &             \\
Ang. dist [arcsec] & 5.43          & 10.3               & 16.1           & 19.0        \\
Proj. Dist. [kpc]  & 8.0           & 62                 &                &
\enddata
\tablecomments{}
\label{tab:Gals}
\end{deluxetable*}

We suggest that there are at least two possibilities
for the identity of the host galaxy of GRB\,211211A.
The first possibility is that, as advocated by \cite{2022lGRB-KN-Rastinejad},
galaxy $A$ is the host galaxy.
With $z=0.076$, this will put the GRB at projected distance of about 8\,kpc from the host, at a position in which no apparent star formation is seen in the HST images.

The deep HST image presented by \cite{2022lGRB-KN-Rastinejad} exclude any underline host to F606W limiting magnitude of 27.8. For a point source, this translate to absolute magnitude of about $-10$ at $z=0.076$.
For an extended source of size of about 1\,kpc it may be possible to hide,
in this position, a brighter galaxy (about $-13$ assuming 1\,kpc size).
The H-band limit reported in \citet{2022lGRB-KN-Troja} is similar, 27.6. However, since the typical expected V-H color of star-forming galaxies is about +1 mag (AB), the H-band limit is somewhat more restrictive. A possible problem with this scenario is that it will be difficult
to a hide a supernova that accompanies the GRB (see \S\ref{sec:SN}).

An alternative is that the GRB is at a higher redshift. For example, a 1 kpc size dwarf at $z\sim0.5$ will be barely resolved by HST, and in this case, in order to hide the host galaxy from detection we require it to be fainter than an absolute magnitude of about $-13.5$ to $-14.5$ ($\sim 0.01 L_{*}$), where the exact limit depends on the size and light distribution of the galaxy.
However, such small galaxy usually resides in massive halos.
Indeed galaxy $C$, $D$ have photometric redshift of about $0.5$
and galaxy $B$ has a spectroscopic redshift of $0.469$.
The chance-coincidence probability to find such a galaxy within
$\cong10''$ from a random position is of order unity.
Therefore, this is not an evidence for an association
of GRB\,211211A with a galaxy group at $z\approx0.5$,
but only a consistency test for this hypothesis.
If indeed GRB\,211211A resides in the halo of galaxy $B$ at $z=0.469$,
than it is located at projected distance of about 62\,kpc from this
galaxy -- A reasonable distance between a dwarf galaxy and its main halo.
Another hint that shows the consistency of this solution
is that the SDSS spectrum of galaxy $B$ display narrow emission lines
indicating of high star formation rate in this environment.

From the information collected by \cite{Blanchard+2016ApJ_LongGRB_Hosts}, we estimate that at least 10\% of the long GRBs have hosts as faint as $0.01 L_{*}$ (see a detailed explanation in appendix \ref{sec:lowLGal-GRB}). Furthermore, these authors demonstrated that luminous galaxies usually associated with long GRBs have a higher chance of coincidence probability. In other words, when looking for the host of a GRB, there is a selection bias towards nearby bright galaxies (which are not the real hosts).

Finally, we note that the upper limit on H$\beta$ emission of about $10\,\mu$Jy from the spectrum of \citet{2022lGRB-KN-Rastinejad} does not set a constraining upper limit on the star-formation rate, that may be expected for a GRB arising from a massive star progenitor residing in a star-forming region. Assuming case-B recombination, 10\,\AA~resolution, $z\approx0.5$, and using the relation between $L$(H$\alpha$) and star formation rate \citep{1998Kennicutt} 
\begin{equation}
SFR({\rm M}_{\odot}\,{\rm yr}^{-1}) = 7.9\times10^{-42} L_{H\alpha}(1+z) {\rm erg}\,{\rm s}^{-1},
\end{equation}
this translates to an upper limit on the star formation rate of about $\lesssim 2 (5)$\,M$_{\odot}$\,yr$^{-1}$ at $z=0.5$ ($z=0.65$).
\cite{2022lGRB-KN-Troja} estimated a somewhat lower upper limit of 1\,M$_{\odot}$\,yr$^{-1}$ at $z=0.65$. While we cannot track the source of the discrepancy between our and their estimate, these limits are not very constraining given the large uncertainties in estimates of this type (e.g., due to extinction; see \cite{1998Kennicutt}),
and the fact that the median estimated star formation rate of GRB-hosting galaxies with $z<1$ is about 1\,M$_{\odot}$\,yr$^{-1}$ \citep{2019GRB-SFR}.

\section{Conclusions}
\label{sec:summary}

\cite{2022lGRB-KN-Rastinejad}, as well as \cite{2022lGRB-KN-Troja} and \cite{2022lGRB-KN-Yang}, argued that the best explanation for the long GRB\,211211A is a kilonova from a NS-compact object merger.
This claim is based on the following arguments:
(i) A host galaxy at 350\,Mpc;
and (ii) Bright NIR signal on time scales of days.
If true, an immediate important consequence is that
NS-merger produces diverse optical signal,
and that they can produce long GRBs.

Following \cite{WaxmanDraine00}, here we suggest another explanation -- GRB\,211211A is
a long GRB in which the NIR excess is the result of thermal emission from dust, heated by UV-soft X-ray radiation produced by the interaction of GRB plasma with the CSM. The heated dust explanation naturally accounts for the large ratio between the NIR and optical specific fluxes, due to the low temperature associated with dust sublimation-- see equations~(\ref{eq:Tc}) and~(\ref{eq:lam_p}), and for the observed NIR luminosity -- see Equation~(\ref{eq:Lratio}). Furthermore, the jet opening angle implied by the observed duration of the NIR emission, Equation~(\ref{eq:thetaD}), is consistent with the opening angle inferred from the break in the X-ray light curve, Equation~(\ref{eq:thetaX}).

The parameters of the gas cloud, that are required to produce the observed NIR luminosity and to avoid strong absorption of the X-ray afterglow, are consistent with those of common molecular cloud values. The cloud radius is required to be comparable to the sublimation radius, $\approx10$\,pc (see Eq.~\ref{eq:Rd}), and the cloud mass is required to be $\lesssim10^{5}\,M_\odot$ (see Eq.~\ref{eq:Mion}). The Hydrogen column density of a $10^{4.5}\,M_\odot$ cloud with $\approx10$\,pc radius is $10^{22}{\rm cm^{-2}}$, a typical value \citep[e.g.][]{1987mol_cloud} corresponding to $\tau_V\approx 4$. For such a cloud a residual $\tau_V\approx1$ is natural (does not require fine tuning of cloud radius or burst location), given that the dust destruction distance is comparable to the cloud's size.

The residual optical depth required for efficient conversion of energy to IR radiation, $0.2\lesssim\tau_V\lesssim2$ corresponding to $0.05\lesssim E(B-V)\lesssim0.5$, may be difficult to identify. The determination of extinction is complicated since it requires an accurate knowledge of the intrinsic source spectrum. This is particularly challenging in the current case, where the intrinsic source spectrum is composed of two independent components, the “afterglow” and red (infra-red dominated) components, both of which are subject to significant uncertainties. The large uncertainty in determining the extinction is reflected, for example, in the large difference between the limits inferred from the early afterglow in the analyses of \citet{2022lGRB-KN-Gompertz}, who obtained $E(B-V)<0.2$, and \citet{2022lGRB-KN-Troja}, who obtained $E(B-V)<0.005$. 

Nevertheless, we note that there may be an indication for a significant residual UV optical depth. \citet{2022lGRB-KN-Gompertz} find a 
significant suppression of the observed optical-UV flux at $t=170$~s compared to the flux expected based on the low-energy extension of the soft X-ray synchrotron flux observed at the same time. They note that this suppression requires a dust optical depth $\tau_V\approx2$ ($E(B-V)\approx 0.5$), but dismiss this explanation based on the argument that such large extinction is inconsistent with the later afterglow spectrum (suggesting that another, yet unknown explanation of the suppression should be found). We point out that this is not a strong argument, as the afterglow spectrum is red and requires the inclusion of a low temperature thermal emission. The inferred dust optical depth $\tau_V\approx2$ corresponds, for typical local clouds, to a H column density of $\approx 5\times10^{21}{\rm cm^{-2}}$ \citep[e.g.][]{Longair92Book,2000WilmsISM-X-ray}. This is larger than the column density inferred by \citet{2022lGRB-KN-Gompertz} (their Table 1) for GRB211211A based on the soft X-ray spectrum, $N_{H,X}\approx 1\times10^{21}{\rm cm^{-2}}$. However, it should be noted that (i) the uncertainty in the inferred column density is large due to the uncertainty in the underlying X-ray spectrum, (ii) the dust properties at the $z=0.5$ host may differ from local cloud dust properties, and (iii) X-ray ionization of metals in the gas phase, which reduces the ionization optical depth and hence the inferred $N_{H,X}$, and fission of dust grains, which may increase the UV extinction (WD00, see also appendix \ref{sec:appendix}), are expected to increase the $\tau_V/N_{H,X}$ ratio.

The dust emission explanation requires the GRB to reside in a galaxy at $z<1$ (see \S~\ref{sec:GRBA}). Our analysis of the GRB\,211211A environment suggests that there are {\it at least} two solutions for the redshift of GRB\,211211A. The first is at $z=0.076$, and the second is at $z\approx0.5$ (see \S~\ref{sec:host}). While both solutions are consistent with the thermal emission from dust scenario, the first solution likely requires a GRB without a supernova and the second solution can accommodate a normal long GRB (with typcial $\gamma$-ray energy) and a supernova (see \S~\ref{sec:SN}). It should be noted that stringent upper limits on the presence of a supernova (100 times fainter than the supernova associated with GRB970508) have been obtained for several long GRBs \citep[e.g.][]{2006DellaValleNoSN,2006Fynbo-NoSN,2006GalYamNoSN}, and do not necessarily imply a compact binary merger progenitor- A "failed supernova" followed by mass accretion onto the newly formed compact object is an alternative explanation. The relatively large, 60~kpc, offset of the burst location from the potential $z\approx0.5$ host implies that, for this solution, the GRB host is likely a dwarf galaxy residing in the halo of the identified massive $z\approx0.5$ galaxy. The upper limit on the luminosity of a host galaxy at the GRB location implies an upper limit of $\approx 0.01 L_\star$ on the dwarf luminosity. This is not unreasonable, as \cite{Blanchard+2016ApJ_LongGRB_Hosts}
estimate that at least 10\% of the long GRBs are hosted in fainter than $0.01 L_{*}$ galaxies. Detecting a putative dwarf host at $z\sim0.5$ may be possible by long integration time with HST and JWST.

To conclude, we provide in Table~\ref{tab:obs_summary} a comparative summary of the compatibility of the key observed properties of GRB 211211A with both the dust heating model and the NS merger kilonova model.
While the dust heating model naturally explains the observed NIR excess and the GRB $\gamma$-ray properties, it struggles with the undetected host galaxy.
The NS merger kilonova model provides a plausible explanation for the host galaxy, but it is challenged by the GRB NIR and $\gamma$-ray properties. This comparison underscores the importance of multi-wavelength observations of future events for distinguishing between these scenarios.

\begin{deluxetable*}{p{3cm}p{4cm}p{4cm}p{4cm}}
\label{tab:obs_summary}
\tablecolumns{4}
\tablewidth{0pt}
\tablecaption{Model consistency with key observed properties of GRB\,211211A}
\tablehead{
\colhead{Observable} &
\colhead{Description} &
\colhead{Dust heating Model, $z\simeq0.5$} &
\colhead{NS Merger kilonova Model, $z=0.08$}}
\startdata
GRB Duration & $\sim50\,\rm{s}$ gamma-ray emission. & Consistent with the typical duration of long GRBs. 
& 
The long duration is a challenge for NS merger driven GRBs.
\\
Distance-dependent GRB $\gamma$-ray properties, \{peak luminosity vs. temporal lag, spectral peak energy vs. isotropic $\gamma$-ray energy\}. & Inconsistent with both the long and short GRB populations for $z = 0.08$ origin, typical to long GRBs for $z\sim0.5$ & Consistent with a long GRB at a $z\sim0.5$ host. & Unusual for both long and short GRBs for a $z=0.08$ host. \\
Strong NIR emission & NIR emission lasting $\sim10\,\rm{d}$, $f_\nu\approx 2\,\mu$Jy at $2.2\,\mu$m. & Consistent with the predicted thermal emission from dust heated by a GRB in (the vicinity of) a molecular cloud.
& Explained as kilonova emission from radioactive decay.\\
Large ratio of NIR to shorter wave-length flux & $f_{\nu,K}/f_{\nu, i}\approx 30$ at $t\approx 5$\,d. & Consistent with thermal dust emission. & Much larger than that observed in GW170817, and difficult to reconcile with existing models for any radioactive ejecta.\\
Host galaxy & Host galaxy candidates at $z=0.076$ and $z\sim0.5$. & 
Requires an undetected, $L<0.01L_{*}$, dwarf in the halo of the $z\sim0.5$ galaxy. & Consistent with a $z=0.076$ host.\\
Dust Extinction Limits & Various model-dependent estimates, e.g. $E(B-V)<0.005$ or $E(B-V)<0.2$. & Requires residual dust extinction of $0.05<E(B-V)<0.5$. & Compatible with minimal host galaxy extinction.\\
Lack of a supernova. & No clear SN signature detected. & Consistent with a relatively dim SN at $z\gtrsim0.5$. & Compatible with a NS merger progenitor.\\
\enddata
\end{deluxetable*}

\begin{acknowledgments}
We thank the anonymous referee for useful comments that led to improvements of the manuscript.
E. W.'s research is partially supported by ISF, GIF and IMOS grants. DK is supported by a research grant from The Abramson Family Center for Young Scientists, by ISF grant, and by the Minerva Stiftung.
E.O.O. is grateful for the support of
grants from the
Norman E Alexander Family M Foundation ULTRASAT Data Center Fund,
Israel Science Foundation,
Israeli Ministry of Science,
Minerva,
NSF-BSF,
Israel Council for Higher Education (VATAT),
Sagol Weizmann-MIT,
and Yeda-Sela.

\end{acknowledgments}

\appendix

\section{The contribution of $>0.1$~keV photons to dust destruction and sublimation}
\label{sec:appendix}

The absorption of $>0.1$~keV photons is dominated by photo-ionization of "metals", i.e. of elements heavier than He. For typical cloud "metal" content, the ionization cross section per H atom is approximately given by $2\times10^{-22}{\rm cm^{-2}}(h\nu/1\,{\rm keV})^{-8/3}$ \citep[e.g.][]{Longair92Book,2000WilmsISM-X-ray}. The rapid decrease of the cross section with energy implies that $h\nu>10$\,keV photons typically deposit only a small fraction of their energy in the cloud. Furthermore, the secondary electrons produced by the ionization of "metals" in dust grains by $h\nu>10$\,keV photons typically escape the grains, leading to grain charging and fission, rather than heating. Grain fission is dominated by $\sim10$\,keV photons since (i) they dominate the number flux of high energy photons, and (ii) the ionization cross section decreases at higher energy. Bright bursts may lead to grain fission out to $\sim10$\,pc (WD00), comparable to the dust sublimation distance, changing the grain size distribution and hence the optical extinction curve. Reducing the grain sizes does not affect significantly the extinction at IR wavelengths, which are large compared to the grain size, while it may increase the extinction at UV wavelengths, which are smaller than or comparable to the grain size. It should be noted that fission of smaller grains requires a larger photon flux, while sublimation of smaller grains requires a lower photon flux. At the sublimation radius $R_d$, fission is typically expected to reduce grain sizes to $\simeq0.1\mu$m (see eqs. 23 and 17 of WD00), and the smaller sizes destruction would be dominated by sublimation.

Let us consider next the contribution of $\sim1$\,keV photons. The luminosity of these photons is typically $\sim10$ times higher than that of non-ionizing $<10$~eV photons (since $\nu L_\nu\propto \nu^{1/2}$ for frequencies below $h\nu_c\gtrsim1$\,keV), their ionization cross section per H atom is similar to the absorption cross section per H atom of optical photons, and the electrons produced by the ionization of "metals" in dust grains by $h\nu\sim1$\,keV photons typically deposit their energy in the grains. One may thus reach the conclusion that $\sim1$\,keV photons dominate grain heating. This is, however, not the case. First, the ionization cross section of $\sim1$\,keV photons is $\lesssim10^{-19}{\rm cm^{-2}}$ for the atoms dominating the ionization \citep[in particular those largely "locked" in grains - C, N, O, Si, S][]{Longair92Book,2000WilmsISM-X-ray}, implying that the optical depth for absorption in a grain is $0.1$ for a $0.1\,\mu$m grain. This is in contrast with the situation for UV photons, which are completely absorbed within a grain's column density. Thus, the heating rate of the grains by $\sim1$\,keV photons is at most comparable to that by non-ionizing photons.
Second, the absorption of $<1$\,keV photons is dominated by ionization of "metals" in the gas phase, rather than by ionization of "metals" contained in grains. For typical cloud "metal" and dust content, only $\approx25\%$ of the ionizations by $\sim1$\,keV photons will take place within grains, while at lower photon energy only a negligible fraction of the ionizations take place within grains (This is due to the "shielding" of most of the grain mass by the outer grain shell absorbing the photons). Absorption in the gas phase strongly suppresses dust heating by $<1$~keV photons. Heating by $\sim1$~keV photons may become comparable to that by non-ionizing photons for bright bursts, or lower mass clouds, where the ionizing photon fluence is sufficient to largely ionize the metals in the gas phase of the cloud.

\section{Faint host galaxies of long GRBs}
\label{sec:lowLGal-GRB}

\citet{Blanchard+2016ApJ_LongGRB_Hosts} searched for the host galaxies of 105 long GRBs. For each candidate host galaxy, they searched for a host galaxy, measured its redshift, and estimated its luminosity and the probability of chance coincidence. 90 GRBs in their sample have a redshift, 79 have a candidate host galaxy, and 72 have both a candidate host and redshift. According to Figure~3 in their paper, out of 72 hosts, 3 events have galaxies with $L<10^{-2}\,L*$. However, 14 events have upper limits which are below $10^{-1}\,L*$. This suggests that even if we ignore the 33 events in their sample which have no candidate host and redshift, we can not rule out the possibility that about 10\% ($<22$\% at a 95\%-confidence level) of the long GRB population originates from $L<10^{-2}\,L*$ hosts. 

Furthermore, from this figure, it is evident that there is an observational bias in the selection process of GRBs host galaxies. Specifically, this figure hints at a correlation between the host galaxy's luminosity and the chance-coincidence probability, showing that higher luminosity galaxies tend to have higher chance-coincidence probability. We find that the Pearson correlation coefficient between the chance-coincident probability ($P_{\rm cc}$) and the host galaxy abs. mag. (not K-corrected) is $-0.20$ with a false alarm probability of 0.04\%. Furthermore, there is a significant correlation between $P_{\rm cc}$ and $z$ ($+0.32$, with a false alarm probability of 0.006).

We conclude that the \citet{Blanchard+2016ApJ_LongGRB_Hosts} sample may contain some wrong host identifications biased towards brighter hosts, and that even if one ignores this bias and all the GRBs with not enough information, this sample is consistent with the fraction of GRB hosts with luminosity smaller than $0.01\,L*$ being of the order of 10\%.

\end{document}